# Advanced spike sorting approaches in implantable VLSI wireless brain computer interfaces: a survey

Soujatya Sarkar
*Department of Biological Sciences*
*Bose Institute, India*
email: soujatyasarkar@jcbose.ac.in

*Abstract*—Brain Computer/Machine Interfaces (BCI/BMIs) have substantial potential for enhancing the lives of disabled individuals by restoring functionalities of missing body parts or allowing paralyzed individuals to regain speech and other motor capabilities. Due to severe health hazards arising from skull incisions required for wired BCI/BMIs, scientists are focusing on developing VLSI wireless BCI implants using biomaterials. However, significant challenges, like power efficiency and implant size, persist in creating reliable and efficient wireless BCI implants. With advanced spike sorting techniques, VLSI wireless BCI implants can function within the power and size constraints while maintaining neural spike classification accuracy. This study explores advanced spike sorting techniques to overcome these hurdles and enable VLSI wireless BCI/BMI implants to transmit data efficiently and achieve high accuracy.

## I. INTRODUCTION

In order for humanity to attain full autonomous control over their biological bodies, a comprehensive exploration into the realm of Brain Computer Interfaces (BCIs) is anticipated. This endeavor aims to uncover the intricacies of our physiological functions and develop advanced technological tools to overcome all forms of disabilities. BCIs hold enormous potential in the quest of granting humans the ability to gain complete autonomy over their biological bodies. This transformational technology entails building a direct and seamless communication channel between the human brain and external technological interfaces. BCIs can potentially improve various areas of human life, notably overcoming impairments and enhancing overall well-being [1], [2].

BCIs provide a unique avenue to deeply understand our physiological functions in unprecedented detail. We have gained profound insights into human mechanisms by deciphering intricate neural patterns and processes administering actions, sensations, and cognition. BCIs bridge the mind-world gap, allowing people with physical disabilities to regain functions through neural signal-controlled devices. This technology promises to restore movement and independence in the impaired masses. BCIs can enable innovative interactions with the environment, such as thought-controlled equipment and prostheses [3]. Finally, BCIs transform the human experience by unlocking huge potential, releasing people from impairments, and promoting autonomy. Despite massive research efforts in the BCI sector, human usage of BCI remains limited due to a lack of safe and reliable invasive VLSI wireless BCI implants [4].

Electroencephalography (EEG) is a non-invasive neural recording method employed to capture the electrical activity of the human brain [5]. Electrocorticography (ECoG) is a semi-invasive neural recording method that involves placing electrodes directly on the brain's surface to capture electrical activity [5]. Alternative approaches, such as magnetic and infrared reasoning, offer advantages in mitigating interference caused by bodily noise artifacts [6]. These alternatives offer improved precision in pinpointing neuron locations and a heightened signal-to-noise ratio (SNR) [7]. Nonetheless, the viability of these methods for implantation and individual everyday usage is constrained due to the substantial size of the required equipment and its associated prerequisites. Hence, deep brain activity recording using intracortical electrodes is considered the most effective and viable method for capturing the brain's neural activity due to its complex instrumentation being integrated as a VLSI chipset, which can be invasively placed inside the brain.

Wireless implants offer the advantage of avoiding chronic skull lesions caused by long-term incisions made for wired BCI systems in the past [8]. However, a problem arises when the implant attempts to transmit every recorded neuron's electric activity because doing so would place substantial pressure on both bandwidth and power, which, in turn, could potentially clash with the strict power constraints of the implant itself, typically falling within the range of 8 to 10 milliwatts [9]. To address this issue, spike sorting methods are harnessed to effectively mitigate the volume of data that needs to be transmitted, thereby alleviating the burden on bandwidth and conserving power resources.

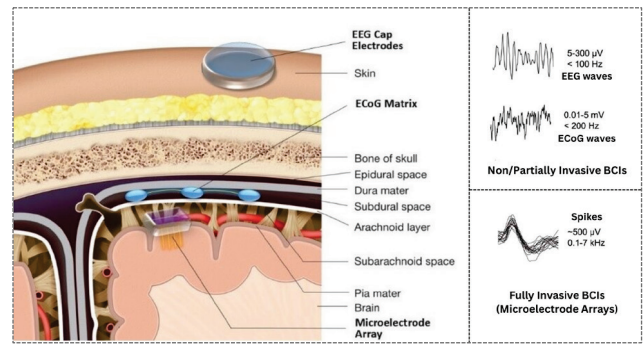

Fig. 1. BCI electrode types and their placement: EEG cap electrodes on the skull (non-invasive), ECoG electrode matrix on the brain surface (partially invasive), and Intracortical microelectrode array penetrating inside the brain (fully invasive). Schematic of the neural signals capturable by the respective electrode types and their properties are described on the right side of the figure. Reproduced with permission [10]. © 2022 IEEE.

This study will introduce us to advanced spike sorting techniques that allow VLSI wireless BCIs to work at lesser bandwidth and power, yield highly accurate results, and become equally or more potent than their wired counterparts.

## II. BCIs: AN OVERVIEW

### A. What are BCIs?

Brain Computer Interfaces (BCIs) are innovative devices facilitating direct communication between the human brain and external devices or computer systems. BCIs tap into the brain's electrical or neural activity, converting it into actionable commands that control various applications [1].

BCIs process captured EEG/ECoG/intracortical electrode wave signals to extract meaningful features, which are used to train AI/ML enhanced models to predict the exact neural commands of the brain accurately. This would enable the intelligent models to mimic specific commands of the brain, which are predicted based on the electric impulses generated, thus allowing them to overcome all kinds of disabilities [11].

*B. Wireless BCIs and its applications*

Typically, the fundamental components of a BCI system are a signal enhancer, feature extractor, clustering algorithm, feature classifier, and post-processing units [4]. This architecture is better suited for software implementation on machines with immense processing power. For interaction with hardware interfaces and robots, a 4-phase wireless BCI architecture is more suited [12], [13].

A wireless BCI initiates by capturing neural signals through intracortical electrodes, like microelectrode arrays, microwires, and multisite probes, which represent brain electrical activity. The obtained neural signals are processed for quality improvement and feature extraction. Signal processing techniques eliminate noise and artifacts to ensure correct interpretation. Extracted features, encompassing frequency bands, and ERPs, symbolize mental states. Machine learning algorithms, trained on labeled data, categorize features—e.g., distinguishing hand movement from focused attention. Commands are wirelessly sent to external devices, often via Bluetooth or Wi-Fi, which interpret user intentions—such as controlling limbs or devices. This real-time process permits immediate actions based on user-generated brain patterns. Some wireless BCIs offer sensory feedback, refining user control. The accuracy of BCIs improves through calibration and adaptation sessions. Wireless BCIs have potential across many industries because they provide comfort, mobility, and real-world applicability. In essence, they capture, process, classify, and wirelessly transmit neural signals to control external systems [5], [14].

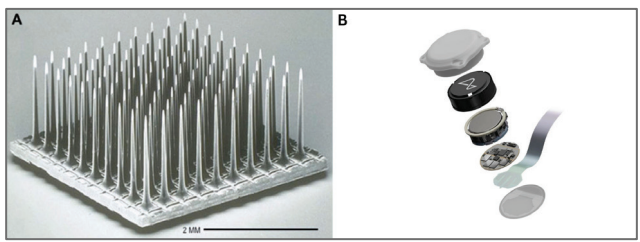

Fig. 2. Popular BCI wireless implants: A) Scanning Electron Micrograph of Utah Microelectrode Array. Reproduced with permission (Licensed under CC BY 3.0) [15]. © 2014 Fernández, Greger, House, Aranda, Botella, Albisua, Soto-Sánchez, Alfaro and Normann, B) Neuralink N1 wireless BCI implant. Reproduced with appropriate citation [16]. © 2023 Neuralink.

The semi-invasive and invasive wired EEG/ECoG-based BCIs are impractical for daily human life as they necessitate routine checkups and raise the risk of infection. A wireless implant solution with an information extraction unit can provide accurate results with minimal power consumption and compact design, allowing for usage as an implant without the requirement for an open skull incision.

*C. Wireless BCI implant standards*

The following characteristics and restrictions must be kept in mind when designing a usable BCI implant: power utilization (8–10 mW), implant size (~1 cm$^2$), signal-to-noise ratio (SNR) (~8.4 dB), and energy dissipation, which must be kept minimal (the temperature of the neighboring tissues can only raise by 1°C), with the total accuracy of the wireless BCI implementation (~80%) [17]–[19].

### III. SPIKE SORTING APPROACHES

Spike sorting is the process of classifying brain activity spikes into clusters according to the similarity of their shapes. They are computational methods that analyze and categorize individual action potentials, or "spikes" recorded from neural electrode arrays [20].

Typically, single or multiple threshold detectors in time and space domains are used for spike sorting (e.g., pick, latency, or sliding window) [4]. Instead of sending the raw electric signal waveforms of neural activities, which further narrows the transmission bandwidth, the addition of spike sorting units in the BCI implant alongside the microelectrode array can reduce power usage and the amount of transmitted data considerably, thus only filtering out the effective and valuable data to transmit through the limited bandwidth [13].

These techniques are crucial in neuroscience research and Brain-Computer Interface (BCI) applications where accurate identification and classification of neural activity are essential. Bashashati et al. [6] and Makeig et al. [12] introduced five different forms of electrophysiological control sources along with their associated functional units. They also discussed the stages and functionality of a BCI system, such as multiple filters, dimensional reduction using Principal Component Analysis (PCA), Independent Component Analysis (ICA), and Data Mining [4]. Most of these methods can be applied in parametrized form in spike sorting algorithms for VLSI-based BCI implants.

*A. PCA-based Spike Sorting Technique*

PCA-based spike sorting is a mathematical method in neuroscience for studying neural activity. It transforms high-dimensional spike waveforms into a lower-dimensional space using Principal Component Analysis (PCA). This entails building a matrix from recorded spike waveforms and calculating the covariance matrix after mean subtraction. Eigenvalue decomposition of the covariance matrix yields principal components and their corresponding eigenvalues. The eigenvectors and eigenvalues obtained from this matrix help in the dimensional reduction of the data, which is done by selecting the top principal components that capture the most variance, thus maintaining the essential information. Clustering algorithms like K-Means are then applied to grouped spikes in the reduced space, assigning them to distinct neurons. This method facilitates identifying the spikes from distinct neurons, providing insights into neural behavior. However, newer and advanced versions have been developed to tackle problems like noise and overlapping spikes, which not only improves the spike sorting accuracy but also provides deeper insights into neural dynamics [21], [22].

An energy-efficient neural model architecture has been proposed by Peng and Xiao [23], which uses a time-multiplexed Analog Front-end (AFE) and an energy-efficient short-range backscattering RF module that consumes 190 μW of power and has an accuracy of ~86.8% [4].

This feat was accomplished by incorporating an initial preparation phase, an Operational Trans-conductance Amplifier (OTA), and a minimal power Analog-to-Digital

Converter (ADC). Additionally, they introduced a spike sorting system based on Principal Component Analysis (PCA) and K-means, which utilized a fixed Euclidean distance threshold [4]. They integrated a First-In-First-Out cache (FIFO) to optimize memory usage, as shown in Fig. 3.

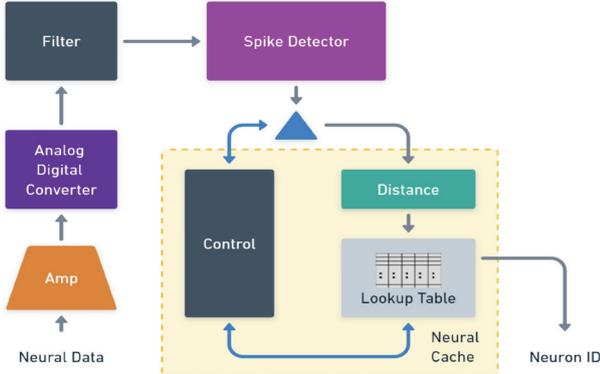

Fig. 3. PCA-based spike sorting system using Neural Cache.

However, a significant issue stems from the concern that the worst-case scenario during the Search and Merge steps might consume O(2n) operations in cases where the data is unsorted. Hence, adopting a cache-hashing technique to ensure a constant time complexity of O(1) is suggested, the only downside being that the enhancement comes with a trade-off involving the amount of memory required and potential power consumed.

B. *Filter-based Spike Detection and Sorting Technique*

Filter-based detection techniques have recently gained considerable attention among other spike-sorting techniques for their simplicity and efficiency in reducing chip area and power dissipation. One such technique is the use of energy-based filters for spike detection. These filters leverage the amplitude and duration of the neural spikes to distinguish them from background noise. A 2008 study by Yang, Chen, and Liu [24] introduced a Digital Signal Processor (DSP) chip designed explicitly for spike sorting that employs energy-based filters for spike detection and discrete-derivative methods for feature extraction. This combination allows for a data rate reduction of over 90%.

Chae, Yang, and Liu [25] put forth one of the foundational ideas for on-chip spike sorting involving the usage of a NEO-based spike detector followed by a series of noise shaping filters. Fig. 4 displays the spike sorting unit.

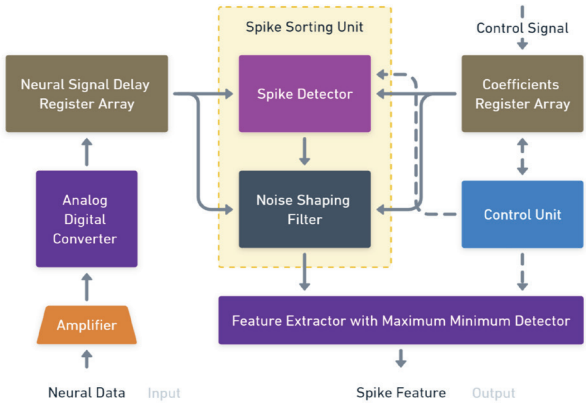

Fig. 4. Filter-based spike sorting unit using Noise Shaping Filter.

To preprocess the raw neural data for the nonlinear energy operator (NEO) spike sorter with a noise shaping filter (a derivative of a standard frequency shaping filter), the system uses a 9-bit Analog-to-Digital Converter (ADC). This not only minimizes the need for memory but also remarkably improves neural recognition across the background noise.

C. *Backboard-based Spike Sorting Technique*

The backboard-based spike sorting technique is a novel approach for accurate neural signal analysis. It focuses on optimizing signal processing using a hierarchical framework. Initially, recorded data is pre-processed to extract spike waveforms. A "backboard matrix" constructed by aligning and concatenating the spike waveforms captures their temporal patterns. Clustering is performed on this matrix utilizing algorithms like K-means, which allow the detection of spike clusters with similar temporal patterns. This technique provides insights into complex neural activities and exploits the temporal relationships between the spikes to enhance the spike sorting accuracy [26], [27].

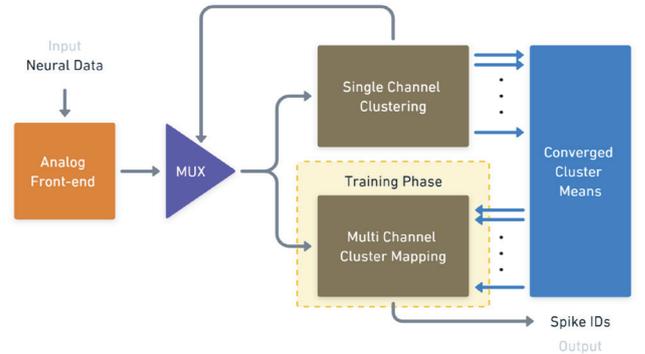

Fig. 5. Schematic diagram of a two-phase online neural spike sorting unit utilizing a "backboard matrix" of converged cluster means.

In addition to the hardware implementable spike sorting solution developed by Gibson, Judy, and Marković [26], Karkare, Gibson, and Marković [27] designed a 16-channel spike sorting Digital Signal Processor (DSP) chipset with a focus on low power consumption and noise tolerance, as depicted in Fig. 5. They innovatively introduced the concept of incorporating spike ID rather than the actual values. The primary algorithm retained its foundation in Analog Front-End (AFE) processing, utilizing the Euclidean distance metric for clustering. This approach integrated an adaptable threshold equation, and to maintain efficiency, it constrained the number of saturated clusters to 16 for every 50 iterations. This reduced the implant size and power usage to a remarkable 75 μW.

Despite the potential minimal power utilization, the main procedure and training phase, which depend on the clustering algorithm, might significantly impair the performance and power utilization in case of excessive channels, raising the required implant size to >1.23 mm$^2$.

D. *Modified K-Means-based Spike Sorting Technique*

Saeed and Kamboh [28] conducted a study on an on-chip spike sorting methodology with an online and unsupervised hardware architecture. Within each stage of the spike sorting process, which involves a spike detector, feature extractor, and classifier, they extensively investigated various methods and selected, modified, and bounded these techniques, namely Teager's Energy Operator (TEO) for Spike Detection, Zero-Crossing Features (ZCF) for Feature Extraction, and Moving

Centroid K-means (MCK) for Classification. Additionally, they introduced an innovative architecture encompassing five classifiers [4]. Fig. 6 visually represents the comprehensive structure of the entire system.

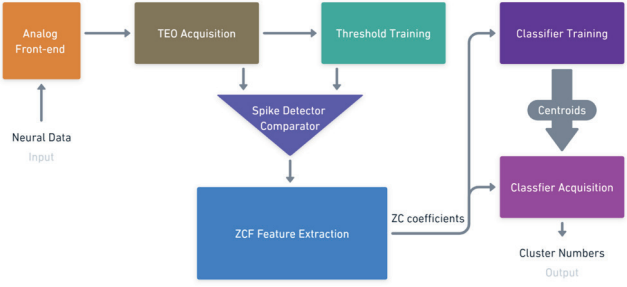

Fig. 6. Schematic diagram of an online neural spike sorting unit based on a modified K-means clustering algorithm.

While it has potential for real-life implementation, the architecture must reduce the training period and bandwidth to comply with implant restrictions.

Conversely, Kamboh and Mason [29] introduced an alternative implementation of an improved Zero-Crossing Features (ZCF) technique. This implementation demonstrated superior performance to PCA-based spike sorting [4] methods, offering an unsupervised heuristic [4] for discrimination without requiring external off-chip training. They further introduced a novel set of computational features tailored for spike sorting, which exhibited a resource utilization of 5%.

### E. Adaptive detection and Skew-t-based Sorting Technique

In 2021, Toosi, Akhaee, and Dehaqani developed an automatic spike sorting technique based on adaptive spike detection and a mixture of skew-t distributions (MSTD) [30]. They have devised an automated spike sorting algorithm that utilizes adaptive spike detection and introduces a mixture of skew-t distributions to mitigate distortions and instabilities. The adaptive detection technique is applied to the detected spikes and involves multi-point alignment and statistical filtering to eliminate erroneously detected spikes. Clustering of the detected spikes is performed based on the mixture of skew-t distributions, addressing non-symmetrical clusters and spike loss issues.

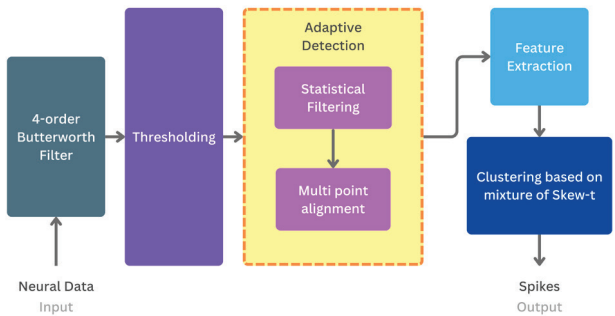

Fig. 7. Schematic diagram of the automatic spike sorting algorithm based on adaptive spike detection and a mixture of skew-t distributions.

The spike sorting algorithm [30] includes three main parts: preprocessing, adaptive detection and clustering. During the preprocessing phase, an initial detection is carried out. To mitigate the impact of undesirable components, a fourth-order Butterworth bandpass filter ranging from 300 to 3000 Hz is applied. Subsequently, in the adaptive detection stage, tasks such as noise removal and multi-point alignment are addressed. Finally, following the feature extraction process, the spikes are automatically clustered using a mixture of skewed multivariate t distributions.

This algorithm [30] claims to enhance spike sorting performance in real datasets, improving precision by ~14%, recall by ~17% and clustering accuracy by ~4% across a wide range of signal-to-noise ratios against its direct predecessor, a regularly used statistical algorithm since its inception in 2003 called the mixture of t-distributions technique (MTD) [31]. Furthermore, the algorithm asserts its validation across ~70% of the diverse datasets having ~90% accuracy, showcasing its efficacy as a versatile solution for accurate spike sorting in both *in vitro* and *in vivo* environments.

### F. Zydeco-style Spike Sorting Technique

In a 2022 research study, ElSayed, Ozer, Elsayed, and Bayoumi developed a Zydeco-style spike sorting [10] technique whose building blocks include: spike fingerprinting unit, global analysis unit, and adaptive matching unit. The core ideology came from the traditional Zydeco music. Zydeco music, characterized by an up-tempo, syncopated style with a robust rhythmic foundation, features the accordion leading the band and a distinctive washboard known as a frottoir serving as a prominent percussive instrument.

The Zydeco-Style architecture [10], as stated earlier, comprises of three key building blocks, each playing a crucial role in the system's functionality. The Spike Fingerprinting Unit generates a precise fingerprint of the firing neurons, using an adaptive threshold spike detector, pivot finder, and fingerprint generator as its integral subunits. The Global Analysis Unit conducts dominant spike analysis by calculating spike delays from surrounding channels and performs spiking frequency analysis for neural identification, producing global-local fingerprints. The Adaptive Matching Unit is responsible for matching current fingerprint values and storing pre-loaded fingerprints in the Fingerprint Lookup Table (FPLT). This unit incorporates an Artificial Immunity System matching unit (AIS) and utilizes 32Kb of SRAM memory to store the initial fingerprint population, and serves as a working space for the matching process. Their proposed architecture underwent simulation and implementation using MATLAB and Verilog, respectively, to validate and verify the model.

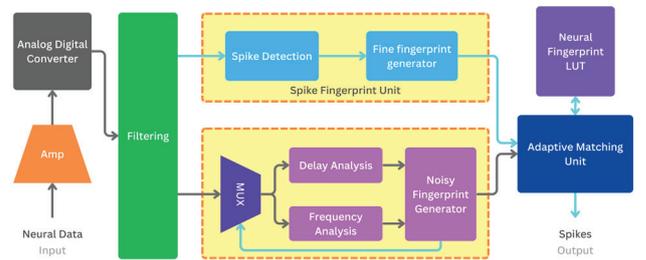

Fig. 8. Schematic diagram of the proposed Zydeco-style spike sorting unit.

The Zydeco-Style architecture [10] exhibits superior computational efficiency and heightened accuracy, achieving up to 93.5% performance even in worst-case scenarios, with optimal detection accuracy observed in the 0dB dataset. However, as SNR exceeds 7dB, system performance accuracy diminishes. Fine-tuning accuracy to ~100% incurs a 34.5% increase in the chip implant area when SNR > 7dB, hence invalidating the model's practical efficacy considering real-world scenarios are non-linear and pose numerous constraints, in contrast to simulated systems.

*G. Neuromorphic model-based Spike Sorting Technique*

In a 2023 research study, Yu, Qi, and Pan [32] developed NeuSort, a novel adaptive spike sorting technique based on neuromorphic models. Utilizing the attributes of neuromorphic models, NeuSort can autonomously acquire the ability to sort spike waveforms and dynamically fine-tune the spike sorting process to accommodate newly incoming neural signals and waveform alterations.

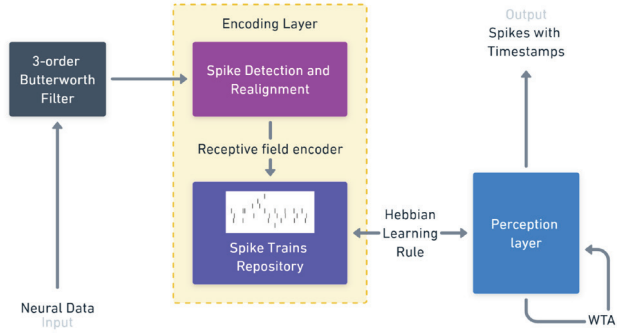

Fig. 9. Schematic diagram of NeuSort spike sorting unit for single channel neural signal data.

In NeuSort [32], the original raw data are converted into spike candidates by following these three steps: bandpass filtering using a 3-order Butterworth filter, NEO-based spike detection, and realignment. These spike candidates are then classified using a receptive field encoder in the encoding layer, which converts their waveforms into spike trains. This encoding layer is connected to a perception layer, where node firing depends on spiking events, and node weights are learned using the Hebbian learning rule. Ultimately, the sorted spikes are output along with timestamps. Fig. 9 describes the framework of the NeuSort spike sorter for a single neural signal channel. In the case of multiple neural signal channels, each channel consists of independent modules following the same spike sorting pipeline, as shown in Fig. 9.

Furthermore, NeuSort has creatively integrated online learning utilizing the Hebbian learning rule, thus allowing it to automatically learn waveform sorting of new-coming neural signals while remembering the previous waveform patterns. While fed with unsorted spike waves continuously, NeuSort progressively develops into an efficient spike sorter by memorizing recurring patterns (neural waveforms) among specific node groups using the synaptic weight update principle of the neuromorphic model, allowing it to trail and adapt to the variations of spike waveforms closely. NeuSort's robust memorization reduces neural signal processing to a single pass, making it the most cost-effective and efficient spike sorting method in terms of bandwidth and power. Its low bandwidth and power requirements allow it to be executable in smaller implant sizes, effectively addressing all the major constraints of VLSI wireless BCIs. The NeuSort algorithm has an accuracy value of ~0.78, a precision value of ~0.88, and a recall value of ~0.86 when compared against real-life test

TABLE I.  SPIKE SORTING TECHNIQUES COMPARISON

| Methods | VLSI Chip | Critical Analysis | | |
|---|---|---|---|---|
| | | *Advantages* | *Challenges* | *Recommendations* |
| PCA-based | 4-channel, 8-bit ADC, 1V, 130nm CMOS, 86.94%. | Unsupervised, power efficient, adaptive, and reduced bandwidth. | The cache can often suffer from slow rebuilding. | Need to improve accuracy and a quick rebuilding cache technique is required. |
| Filter-based | 128-channel, 90Mbit/s, 1.65V, 0.35μm CMOS, 8.8 x 7.2 mm$^2$, 6mW. | Unsupervised, large channel count, accurate, fast, and low power. | Needs preloaded samples. Multiple amplifiers are used. | Diverse data source profiling required. |
| Backboard-based | 16-channel, 0.27V, 65nm CMOS, 1.86 x 1.32 mm$^2$, 75μW. | Unsupervised double-stage clustering, low power and size, provides Spike IDs and uses human neural data. | Requires a training phase. Mainly depends on the number of clusters targeted and wide accuracy variance. | Extra size and accuracy are required to fit this chip into an implant with a wireless module. |
| K-Means-based | MATLAB Simulation. | Unsupervised, adaptive, scalable, online, and improved clustering algorithm. | Requires a training phase and has average accuracy. | VLSI chip design and implantation should be done with respect to power and area. |
| Adaptive spike detection and MSTD-based | MATLAB Simulation. | Unsupervised, adaptive, fast, accurate, and power efficient. | Statistical filter-based. The authors neither discussed the accuracy nor the results properly. | Accuracy analysis and diverse data source profiling required. |
| Zydeco-style | MATLAB and Verilog Simulation. | Unsupervised, large channel count, accurate, spike fingerprinting, fast, and power-efficient. | Suitable spike sorting can only be achieved when SNR = 0dB. When SNR > 7dB, the model requires a larger chip area (~34.5% times). | The chip area constraints need to be maintained to ~1cm$^2$. |
| Neuromorphic-based | MATLAB simulation, Utah 96-channel microelectrode array. | Neuromorphic, large channel count, accurate, fast, online, and power efficient. | Has potential. No major challenges were spotted but accuracy needs to be higher. | Need to improve accuracy to 90% and above. |

datasets, making it the best spike sorting algorithm amongst all other contemporaries.

## IV. Discussion

After studying several types of spike sorting techniques, we can conclude that NeuSort [32], the neuromorphic model-based spike sorting technique, shows great promise and potential amongst all the others in fulfilling humanity's goal of reaching a handicap-free future. Due to the strong memorization power of this algorithm, all neural signals need to pass through the sorter only once, thus making it the cheapest and most efficient spike sorting technique in terms of bandwidth and power consumption among all the others discussed. As the bandwidth and power requirements of NeuSort are very cheap, the expected implant size should be considerably smaller than other implants, thus allowing us to get the best solution for dealing with both of the major concerns of a wireless BCI implant.

Though the modified K-means-based spike sorting techniques showed quite promise, it failed miserably in terms of accuracy and implant size optimization due to its colossal resource hunger regarding bandwidth and memory [28], [29]. The only method discussed that was practically feasible before the discovery of NeuSort is the Filter-based spike sorting technique devised by Chae, Yang, Chen, and Liu collectively [24], [25], which had sufficient accuracy and met the implant size, bandwidth, and power constraints. The backboard-based spike sorting technique also had implant size issues due to the addition of more hardware components [27]. Both the adaptive spike detection with mixture of skew-t distributions [30] and the Zydeco-style [10] techniques have only been simulated in MATLAB, thus not providing enough outlook on how they can be practically used with real-life implants. As the precursor of all spike sorting methodologies, the PCA-based technique has significantly influenced the domain of BCIs and pioneered the notion of wireless VLSI implants for fully invasive BCIs [23].

Thus, NeuSort stands out as a highly promising spike sorting algorithm due to its impressive accuracy, precision, and recall capabilities compared to other algorithms, though it is feasible in smaller implant sizes, lesser bandwidth, and lesser power.

## V. Conclusion

In conclusion, advanced spike sorting techniques play a crucial role in the realm of wireless Brain Computer Interfaces (BCI). The convergence of neuroscientific insights and technological advancements has ushered in a new era of efficient and accurate neural signal processing, enabling real-time applications in wireless BCIs. The ongoing refinement of these techniques aligns seamlessly with the evolution of wireless BCIs, establishing a synergistic relationship between neuroscience and engineering. This evolution holds the promise of transformative impacts on medical treatments, assistive technologies, and our understanding of the intricacies of the human brain through wireless interfaces. We can create efficient and precise VLSI chip implants by integrating spike sorting, clustering, and feature extraction methods with adaptive, parametrized, and online machine learning. While researchers have made strides in designing Spike Sorting DSP chips, there's still substantial work ahead to develop practical wireless BCI implants for human use.

ACKNOWLEDGMENT

I want to thank Prof. Dibyendu Bikash Seal (University of Calcutta, India) and Dr. Priyanka Bhadra (Bose Institute, India) for reviewing, proofreading, and providing valuable insights on this article.